\documentstyle[12pt]{article}
\begin{document}

\thispagestyle{empty}

{}~ \hfill TECHNION-PH-96-10\\
\vspace{1cm}

\begin{center}
{\large \bf  The Long Distance Contribution to $D\to\pi l^+l^-$ }\\
\vspace{1.50cm}
{\bf Paul Singer and Da-Xin Zhang}\\
\vspace{0.50cm}
Department of Physics, Technion -- Israel Institute 
of Technology, Haifa 32000,
Israel\\
\vspace{1.5cm}
{\bf Abstract}
\end{center}

\vspace{1cm}
\noindent 
We calculate the long distance contribution to  $D^{+,0}\to\pi^{+,0}l^+l^-$
decays by the use of a vector meson dominance model,
in which the $\phi$-meson plays the central role.
The branching ratios obtained are $10^{-6}$ and a few times
$10^{-7}$ for the resonance and non-resonance regions respectively.
The analysis includes a  calculation
of $D^+\to\pi^+\phi$, 
consistent with the experimental value.

\vspace{1.2cm}
\noindent {\rm PACS number(s):}  13.20.Fc 12.40.Vv 12.15.Lk

\newpage
\section{INTRODUCTION AND THE MODEL}
New experimental limits for transitions involving change of flavour 
in charm meson decays were obtained recently at FERMILAB\cite{1,2}
and at the Cornell Electron Storage Ring using the CLEO II detector\cite{3}.
In particular, upper limits in the range $10^{-4}-10^{-5}$ were established
for exclusive channels of the type $D^{+,0}\to X^{+,0}l^+l^-$,
where $X^{+,0}$ is a pseudoscalar or vector meson.

The short distance process $c\to u\gamma$, 
driven by the magnetic penguin,
is known to be of little significance for radiative decays of charm mesons
as it leads to a branching ratio of $10^{-12}-10^{-11}$ only,
despite its enhancement by gluonic corrections\cite{4}.
In decays to lepton pairs,
the short distance $c\to ul^+l^-$ transition may be of more relevance,
as it contains contributions from both form factors of the electromagnetic
penguin,
as well as contributions from the $Z^0$-penguin and from the $W$-box
diagram.
Indeed, the rate for the short distance $c\to ul^+l^-$ 
transition has been calculated\cite{5} to be $1.1\times 10^{-20}$GeV,
leading to a branching ratio of $1.8\times 10^{-8}$ for the inclusive process.
Accordingly, exclusive process like $D\to \pi l^+l^-$,
expected to be approximately $10\%$ of the inclusive rate,
will have a short distance contribution to the branching ratio of the order
of $10^{-9}$.
On this basis, it has been argued\cite{6} that decays like 
$D^{+,0}\to\pi^{+,0}l^+l^-$, $D^{+}_s\to K^{+}l^+l^-$ constitute
``... a large ``discovery'' window:
seeing this decay occur at a branching ratio above $\sim 10^{-7}$ 
would be strong evidence for new physics''.

It is therefore of obvious interest to obtain reliable estimates for the 
long distance contributions to these modes.
As one suspects\cite{7} that these contributions are the dominant ones
in the Standard Model,
one should have good control of their  estimation,
in order to perform a meaningful search for new physics in these decays.

In the present paper,
we consider the long distance contribution to the helicity-unsuppressed
decays $D\to \pi l^+l^-$,
which supposedly are the better suited channel for checking the nature of the 
flavour changing neutral transitions.
We note that presently the best experimental upper bounds for the 
branching ratios are as follows:
$1.8\times 10^{-5}$ for $D^+\to \pi^+ \mu^+\mu^-$,
$6.6\times 10^{-5}$ for $D^+\to \pi^+e^+e^-$\cite{2}, and
$1.8\times 10^{-4}$ for $D^0\to \pi^0  \mu^+\mu^-$\cite{1},
$4.5\times 10^{-5}$ for $D^0\to \pi^0 e^+e^-$\cite{3}.

Our estimate for the long distance contribution is based on a vector meson
dominance mechanism,
similar to the approach widely used for calculating the long 
distance contributions in $b\to sl^+l^-$ decay\cite{8}.
We present the details of our calculation for the $D^+\to\pi^+l^+l^-$ channel,
for which the experimental input required by our model is available.

The basic assumption of our model is that the main  long distance contribution 
to $D^+\to\pi^+l^+l^-$ is given by the transition $D^+\to\pi^+(V)\to\pi^+l^+l^-$,
where $V$ is a $\bar{q}q$ vector meson state.
It is known that\cite{9}

\begin{equation}
\begin{array}{rcl}
BR(D^+\to\pi^+\phi)&=&(6.1\pm 0.6)\times 10^{-3},\\
BR(D^+\to\pi^+\rho)&<& 1.4\times 10^{-3},\\
BR(D^+\to\pi^+\omega)&<&7\times 10^{-3}.
\end{array}
\label{e1}
\end{equation}
Moreover, the branching ratios of $\phi$ for decays to lepton pairs are\cite{9}
$BR(\phi\to e^+e^-)=(3.00\pm 0.06)\times 10^{-4}$,
$BR(\phi\to \mu^+\mu^-)=(2.48\pm 0.34)\times 10^{-4}$,
while for $\rho$ and $\omega$ the similar branching ratios
are nearly one order of magnitude lower.
Thus, we may restrict our considerations to the role of the 
$\phi$-meson only,
the contribution of $\rho$ and $\omega$ constituting a relatively small correction.

Another kind of possible long distance contributions comes
from the W-annihilation (or W-exchange) diagram,
which is found for example to be large in 
$B\to \rho\gamma$\cite{ALIWYLER},
and needs therefore to be discussed further.
In the present case of $D\to \pi e^+e^-$,
the mechanism for the  W-annihilation (or W-exchange) diagram
is through $D\to \pi \gamma^*$ and $\gamma^*\to e^+e^-$,
where the virtual photon comes from one of the four quark lines.
Because naive (constituent) quark model  does not work
for the light pseudo-Goldstone meson $\pi$,
we need to use a hadronic model to estimate these contributions.
At the hadronic level,
the process  $D\to \pi \gamma^*$  corresponds to
the electromagnetic transition $D\to D\gamma^*$ followed by
the weak transition $D\to \pi$,
or  $D\to \pi$ followed by $\pi\to \pi \gamma^*$.
These electromagnetic transitions are described
by the electromagnetic formfactors of the $D$ and the $\pi$ mesons,
which can be taken in the VMD model as dominated by   vector
mesons as $\psi$, $\rho$ and $\omega$.
The contribution from $\psi$ is smaller than those from the
other two since this $\psi$ is highly off-shell.
The remaining contributions from $\rho$ and $\omega$
in the W-annihilation (or W-exchange) mechanism
correspond to the full processes $D\to \pi\rho$ and $D\to\pi\omega$.
These effects have been included in our VMD
model and found to be small.

In the second section we present our approach to the
calculation of $D^+\to\pi^+\phi$ and in the third section we calculate its contribution
to the $D\to\pi l^+l^-$ modes.

\section{THE $D^+\to\pi^+\phi$ TRANSITION}

We begin with a treatment of the process $D^+\to\pi^+\phi$,
which is the main contribution to  $D^+\to\pi^+ l^+l^-$ in our approach.
The effective Hamiltonian for this transition is

\begin{equation}
{\cal H}_{eff} = 
\displaystyle\frac{G_F}{\sqrt{2}}
V_{us}^*V_{cs} a_2 \bar{s}\gamma^\mu(1-\gamma_5) s
\bar{u}\gamma_\mu(1-\gamma_5) c,
\label{e2}
\end{equation}
where $a_2$ is QCD-coefficient which is taken as $|a_2|=0.55\pm 0.1$
from the overall fit to nonleptonic $D$ decays\cite{10}.
We shall assume the usefulness of the factorization hypothesis\cite{10},
thus

\begin{equation}
<\pi^+\phi|{\cal H}_{eff}|D^+ >=\displaystyle\frac{G_F}{\sqrt{2}}
V_{us}^*V_{cs} a_2 
<\phi|\bar{s}\gamma^\mu(1-\gamma_5) s |0>
<\pi^+| \bar{u}\gamma_\mu(1-\gamma_5) c|D^+ >,
\label{e3}
\end{equation}
and we define $g_\phi$ by

\begin{equation}
<\phi|\bar{s}\gamma^\mu s|0>=i g_\phi\epsilon^{*\mu}_\phi.
\label{e4}
\end{equation}
$g_\phi$ will be determined from the observed lepton pair decay 
$\phi\to e^+e^-$.

The hadronic matrix element is parameterized by two independent 
form factors $f_+$ and $f_-$ as

\begin{equation}
<\pi^+| \bar{u}\gamma_\mu(1-\gamma_5) c|D^+ >=
f_+(p_D+p_\pi)_\mu +f_-(p_D-p_\pi)_\mu.
\label{e5}
\end{equation}
The $f_-$ form factor does not contribute in the decay to a lepton pair;
$f_+$ is related by isospin symmetry to the form factor of the semileptonic
decay  $D^+\to\pi^0 l^+\bar{\nu}_l$.
Accordingly, we turn to the latter for learning the $f_+$ form factor.

Using the heavy meson chiral perturbation theory, Wise\cite{11} has calculated
the $f_+$ form factor near the zero recoil point 
$q_m^2=(p_D-p_\pi)_m=(m_{D^+}-m_{\pi^0})^2$ in $D^+\to\pi^0 l^+\bar{\nu}_l$
to be:

\begin{equation}
\begin{array}{rcl}
f_+(q_m^2) &=&-\displaystyle\frac{f_D}{2\sqrt{2}f_\pi}
(1-g\displaystyle\frac{p_D\cdot p_\pi-m_D^2}{p_D\cdot p_\pi+m_D\Delta})\\
&\sim&-\displaystyle\frac{f_D}{2\sqrt{2}f_\pi}(1+
g\displaystyle\frac{m_D-m_\pi}{m_\pi+\Delta}),\\
&&\Delta\equiv m_{D^*}-m_D.
\end{array}
\label{e6}
\end{equation}
The analysis of various existing experiments gives the following bounds on the 
parameters in Eq. (\ref{e6})\cite{9,12}:

\begin{equation}
\begin{array}{rcl}
f_D&\leq& 0.31{\rm GeV},\\
|g|&\leq&0.63.
\label{e7}
\end{array}
\end{equation}
On the other hand, two measurements\cite{13} of the formfactor $f_+(q^2)$
in the semileptonic decay have been analyzed under the assumption of a 
monopole behaviour for it,

\begin{equation}
f_+(q^2)=f_+(q_m^2)\displaystyle
\frac{1-q_m^2/m_{D^*}^2}{1-q^2/m_{D^*}^2}
\label{e8}
\end{equation}
and the data may be summarized as\cite{9}

\begin{equation}
f_+(0)=(1.0\pm 0.3)\times f_+^K(0)=(1.0\pm 0.3)\times(0.75 \pm 0.03),
\label{e9}
\end{equation}
which shows the consistency of Eqs. (\ref{e6}) - (\ref{e8}),
within the accuracy obtained so far.
A new recent measurement by E687 gave 
$|f_+^\pi(0)/f_+^K(0)|= 1.03\pm 0.16\pm 0.02$\cite{13a},
which is consistent with the value cited in Eq. (9).

Using now, for exemple, the reasonable set of values $f_D=0.20$GeV and  $g=0.6$,
as well as Eq. (\ref{e8}), we find 
$\Gamma(D^+\to \pi^+\phi)=(3.7\pm 1.4)\times 10^{-15}$GeV,
which is in remarkable agreement with the experimental average\cite{9}
$^{exp}\Gamma(D^+\to \pi^+\phi)=(3.8\pm 0.4)\times 10^{-15}$GeV.
This approach leads to $f_+(0)=0.92$ which agrees with the results of 
Refs. \cite{13,13a}.

In the calculation of the leptonic decay $D^+\to\pi^+ l^+l^-$
we shall normalize to the observed $D^+\to\pi^+\phi$ rate.
Nevertheless, it is appropriate to emphasize at this point that 
the theoretical treatment described here accounts well for this mode;
this provides the needed confidence in its use as the major contribution to the 
$D\to\pi l^+l^-$ decays.

\section{LONG DISTANCE CONTRIBUTION TO  $D^+\to\pi^+ l^+l^-$}

The effective coupling between $\phi$ and $l^+l^-$, via a photon propagator is

\begin{equation}
\begin{array}{rcl}
g_{\phi l\bar l}&=&[e\bar u_{l^-}\gamma^\alpha v_{l^+}]
\displaystyle\frac{g_{\nu\alpha}}{(p_++p_-)^2}
[\displaystyle\frac{1}{3}eg_\phi\epsilon_\phi^\nu]\\
&=&
\displaystyle\frac{e^2}{3}\bar u_{l^-}\gamma^\nu v_{l^+}
\displaystyle\frac{g_\phi}{(p_++p_-)^2}\epsilon_\phi^\nu.
\end{array}
\label{e10}
\end{equation}
For the $\phi$-meson decaying onshell, we replace 
$1/{(p_++p_-)^2}\to 1/{m_\phi^2}$.

We shall assume that $g_\phi(q^2)$ defined in Eq. (\ref{e4})
does not vary appreciably with $q^2$ in the region of interest 
for our calculation\cite{14},
which is taken as $m_\eta\leq\sqrt{m_{+-}^2}\leq (m_D-m_\pi)$,
and we use for it the value determined from $\phi\to e^+e^-$ decay

\begin{equation}
g_\phi=(492{\rm MeV})^2.
\label{e11}
\end{equation}
The amplitude for the decay is then
\begin{equation}
\begin{array}{rl}
&{\cal A}(D^+\to\pi^+l^+l^-)\\
=&
\displaystyle\frac{G_F}{\sqrt{2}}a_2V_{us}^*V_{cs}
<\pi^+| \bar{u}\gamma_\mu(1-\gamma_5) c|D^+ >\\
&
g_\phi \displaystyle\frac{g_{\mu\nu}-
(p_++p_-)_\mu(p_++p_-)_\nu/m_\phi^2}
{(p_++p_-)^2-m_\phi^2+i\Gamma_\phi m_\phi}
\displaystyle\frac{e^2}{3}\bar u_{l^-}\gamma^\nu v_{l^+}
\displaystyle\frac{g_\phi}{(p_++p_-)^2}
\\=&
\displaystyle\frac{G_F}{\sqrt{2}}a_2V_{us}^*V_{cs}
g_\phi^2\displaystyle\frac{2e^2}{3}
\displaystyle\frac{1}{m_{+-}^2}\displaystyle
\frac{1}{m_{+-}^2-m_\phi^2+i\Gamma_\phi m_\phi}
f_+(m_{+-}^2)\bar u_{l^-}{\not\!\! p_\pi} v_{l^+},
\end{array}
\label{e12}
\end{equation}
where we used for the $\phi$ propagator a Breit-Wigner 
form to account for the behaviour throughout the region of decay.

Since at the $\eta$-mass the sizable $D^+\to\pi^+(\eta)\to\pi^+\mu^+\mu^-$
opens [ $BR(D^+\to\pi^+\eta)=(7.5\pm 2.5)\times 10^{-3}$ ],
we shall impose a lower cut on the $m_{+-}$ spectrum of the muon
pairs above the $\eta$ mass.
We have also checked that changing this cut up to $700$MeV does not affect
practically our results.
We shall make for convenience the same cut in the $e^+e^-$ channel. 
Then, the spectra for $e^+e^-$ and $\mu^+\mu^-$ are essentially identical in the 
chosen region for $m_{+-}$.
By restricting our considerations to this region,
we also avoid a possible ambiguity concerning th $q^2$ dependence
in the  region close to $q^2=0$.

The contribution of $\phi$ to the long distance 
$D^+\to\pi^+ l^+l^-$ decay in the resonance region is given by

\begin{equation}
\Gamma_R=
\int_{(m_\phi-\Gamma_\phi/2)^2}^{(m_\phi+\Gamma_\phi/2)^2}
dm_{+-}^2\displaystyle\frac{d\Gamma(D^+\to\pi^+l^+l^-)}{dm_{+-}^2},
\label{e13}
\end{equation}
and we find it to contribute to the branching ratio
\begin{equation}
^{(\Gamma_R)}BR(D^+\to\pi^+l^+l^-)=0.82\times 10^{-6}.
\label{e14}
\end{equation}
If we extend the limits in (\ref{e13}) to $m_\phi\pm \Gamma_\phi$,
this branching ratio becomes $1.22\times 10^{-6}$.

Now we turn to the region outside the resonance,
which is the main object of the present investigation. 
We denote this contribution as $\Gamma_{NR}$, 
and we calculate it from the amplitude of Eq. (\ref{e12}) for the region
$m_\eta\leq m_{+-}\leq (m_\phi-\Delta)$ and $(m_\phi+\Delta)\leq m_{+-}\leq (m_D-m_\pi)$.
The branching ratios for the long distance contribution outside the resonance region,
for several values of $\Delta$, are given in Table 1.

\noindent {\bf TABLE 1.} The long distance contribution $\Gamma_{NR}$
to the $D^+\to \pi^+\mu^+\mu^-$ or $D^+\to \pi^+e^+e^-$ decay rates,
outside the $\phi$-resonance (beyond $m_\phi\pm \Delta$).\\[3mm]
\begin{tabular}{ccc} 
\hline\hline
$\Delta$(in MeV) & $\Gamma_{NR}$ (in $10^{-19}$GeV) &  Branching ratio\\
5& 3.66& $5.9\times 10^{-7}$\\
10&1.92& $3.1\times 10^{-7}$\\
20&0.95&$1.5\times 10^{-7}$\\
40&0.45&$0.73\times 10^{-7}$\\
\hline\hline
\end{tabular}

\vskip 0.3cm

We remind the reader now that the short distance contribution\cite{5} to these decays
is about two orders of magnitude smaller than the values of Tables 1.
Hence, the typical interference distribution in $m_{+-}$ expected in $b\to s l^+l^-$\cite{4}
should not appear in $D\to \pi l^+l^-$,
the spectrum of the lepton pair in our case being given by the matrix element of Eq. (\ref{e12}) and
phase space only.

In the present paper, we concentrated on the  $D^+\to \pi^+l^+l^-$ modes.
For the parallel decays $D^0\to \pi^0l^+l^-$,
we expect in our model a branching ratio smaller by approximately $5$,
due to a factor of $2$ from the $\Delta I=\frac{1}{2}$ weak  $D\to \pi$ transition and
a factor of $2.5$ from the $D^+ - D^0$ lifetime difference.

We remark that a previous long-distance calculation
of $D\to \pi l^+\l^-$ (first Ref. of \cite{7}) has considered it as
evolving from the $D\to D^*\pi$ process,
with the virtual $D^*$  decaying to a lepton pair.
A rate of $10^{-8}$ has been found for this contribution;
this is consistent with our result as this specific diagram is
obviously only part of the form factor we considered here
for the transition (5).

We conclude by stressing that our results,
presented in Eq. (\ref{e13}) and in Table 1,
invalidate the earlier expectations\cite{6} that 
branching ratios above $\sim 10^{-7}$ would constitute a signal for
physics beyond the Standard Model.

This research  was supported in part by Grant 5421-3-96
from the Ministry of
Science and the Arts of Israel. The work
of   P.S. has  been supported in part also by the Fund for Promotion of
Research at the Technion.
One of us (P.S.) acknowledges a helpful discussion with
Professor Lalit Sehgal.
We also thank our colleagues
Prof. Gad Eilam, Dr. Dan Pirjol and Prof. Daniel Wyler
for helpful remarks.


\end{document}